# Diffractive all-optical computing for quantitative phase imaging

Deniz Mengu[a,b,c] and Aydogan Ozcan*[a,b,c]

[a]*Department of Electrical & Computer Engineering, University of California Los Angeles (UCLA), California, USA*

[b]*Department of Bioengineering, University of California Los Angeles (UCLA), California, USA*

[c]*California NanoSystems Institute (CNSI), University of California Los Angeles (UCLA), California, USA*

*\*E-mail: ozcan@ucla.edu*

**Abstract**

Quantitative phase imaging (QPI) is a label-free computational imaging technique that provides optical path length information of specimens. In modern implementations, the quantitative phase image of an object is reconstructed digitally through numerical methods running in a computer, often using iterative algorithms. Here, we demonstrate a diffractive QPI network that can synthesize the quantitative phase image of an object by converting the input phase information of a scene into intensity variations at the output plane. A diffractive QPI network is a specialized all-optical processor designed to perform a quantitative phase-to-intensity transformation through passive diffractive surfaces that are spatially engineered using deep learning and image data. Forming a compact, all-optical network that axially extends only ~200-300λ, where λ is the illumination wavelength, this framework can replace traditional QPI systems and related digital computational burden with a set of passive transmissive layers. All-optical diffractive QPI networks can potentially enable power-efficient, high frame-rate and compact phase imaging systems that might be useful for various applications, including, e.g., on-chip microscopy and sensing.

**Introduction**

Optical imaging of weakly scattering phase objects has been of significant interest for decades, resulting in numerous applications in different fields. For example, the optical examination of cells and tissue samples is frequently used in biological research and medical applications, including disease diagnosis. However, in terms of their optical properties, isolated cells and thin tissue sections (before staining) can be classified as weakly scattering, transparent objects [1]. Hence, when they interact with the incident light in an optical imaging system, the amount of light scattered due to the spatial inhomogeneity of the refractive index is much smaller than the light directly passing through, resulting in a poor image contrast at the output intensity pattern. One way to circumvent this limitation is to convert such phase objects into amplitude-modulated samples using chemical stains or tags [2]. In fact, for over a century, histopathology practice has relied on the staining of biological samples for medical diagnosis to bring contrast to various features of the specimen. While these methods generally provide high-contrast imaging (sometimes with molecular specificity), they are tedious and costly to perform, often involving toxic chemicals and lengthy manual staining procedures. Moreover, the use of exogenous stains might cause changes in the physiology of living cells and tissue, creating practical limitations in various biological applications [3].

The phase contrast imaging principle, invented by Frits Zernike, represents a breakthrough (leading to the 1953 Nobel Prize in Physics) on imaging the intrinsic optical phase delay induced by transparent, phase objects without using exogenous agents [4]. Nomarski's differential interference contrast (DIC) microscopy is another method frequently used to investigate phase objects without staining [5]. While both phase contrast imaging and DIC microscopy can offer sensitivity to nanoscale optical path length variations, they reveal the phase information of the specimen in a *qualitative* manner. On the other hand, quantification and mapping of a sample's phase shift information with high sensitivity and resolution allows for various biomedical applications [6–8]. To address this broad need, quantitative phase imaging (QPI) has emerged as a powerful, label-free approach for optical examination of, e.g., morphology and spatiotemporal dynamics of transparent specimens [3]. The last decades have witnessed the development of numerous digital QPI methods, e.g., Fourier Phase Microscopy (FPM) [9], Hilbert Phase Microscopy (HPM) [10], Digital Holographic Microscopy (DHM) [11–16], Quadriwave Lateral

Shearing Interferometer (QLSI) [17] and many others [18–27]. This transformative progress in QPI methods has fostered various applications in, e.g., pathology [12], cell migration dynamics [6,28] and growth [29], immunology [30] and cancer prognosis [31–34], among others [35–42].

A QPI system, in general, consists of an optical imaging instrument based on conventional components such as lenses, beamsplitters, as well as a computer to run the image reconstruction algorithm that recovers the object phase function from the recorded interferometric measurements. In recent years, QPI methods have also benefited from the ongoing advances in machine learning and GPU-based computing to improve their digital reconstruction speed and spatiotemporal throughput [43–48]. For example, it has been shown that feedforward deep neural networks can be used for solving challenging inverse problems in QPI systems, including, e.g., phase retrieval [49,50], pixel super-resolution [51] and extension of the depth-of-field [52].

In this work, we report the numerical design of diffractive optical networks [53] to replace digital image reconstruction algorithms used in QPI systems with a series of passive optical modulation surfaces that are spatially engineered using deep learning. The presented QPI diffractive networks (Fig. 1) have a compact footprint that axially spans ~240$\lambda$ and are designed using deep learning to encode the optical path length induced by a given input phase object into an output intensity distribution that all-optically reveals the corresponding QPI information of the sample. Through numerical simulations, we show that these QPI diffractive network designs can generalize not only to unseen, new phase images that statistically resemble the training image dataset, but also generalize to entirely new datasets with different object features.

It is important to emphasize that these QPI diffractive networks do *not* perform phase recovery from an intensity measurement or a hologram. In fact, the input information is the phase object itself, and the QPI network is trained to convert this phase information of the input scene into an intensity distribution at the output plane; this way, the normalized output intensity image directly reveals the quantitative phase image of the sample in radians.

The diffractive QPI designs reported in this work represent proof-of-concept demonstrations of a new phase imaging concept, and we believe that such diffractive computational phase imagers can find various applications in on-chip microscopy and sensing due to their compact footprint, all-optical computation speed and low-power operation.

**Results**

Revealing the optical phase delay induced by an input object by converting or encoding the sample information into an optical intensity pattern at the output plane is a relatively old and well-known technique [4]. Unlike analog phase contrast imaging methods that allow qualitative investigation of the samples, modern QPI systems numerically retrieve the spatial map of the optical phase delay induced by the sample. However, the fundamental idea of encoding the phase information of the object function into the output intensity pattern prevails. For instance, coherent QPI methods use optical hardware, commonly based on conventional optical components such as lenses and beamsplitters, to generate interference between a reference wave and the object wave over an image sensor-array, creating fringe patterns that implicitly describe the phase function of the input sample. These QPI systems also rely on a phase recovery step implemented in a computer that decodes the object phase information by digitally processing the recorded optical intensity pattern(s), often using iterative algorithms.

To create an all-optical QPI solution without any digital phase reconstruction algorithm, we designed diffractive networks [53–55] that transform the phase information of the input sample into an output intensity pattern, quantitatively revealing the object phase distribution through an intensity recording. Figure 1 illustrates the schematic of a 5-layer diffractive network that was trained to all-optically synthesize the QPI signal of a given input phase object (see Methods section for training details). This system can precisely quantify and map the optical path length variations at the input, and unlike the modern QPI systems, it does not rely on a computationally intensive phase reconstruction algorithm or a digital computer.

For a proof-of-concept demonstration, here we considered the design of diffractive QPI networks with unit magnification, such that the input object features in the phase space have the same scale as the output intensity features behind the diffractive network. Since the value of the output optical intensity will depend on external physical factors such as, e.g., the power of the illumination source and the quantum efficiency of the image sensor-array, we used a background region (see Methods section) that surrounds the unit magnification output image to obtain a reference mean intensity.

This mean signal intensity value at this background region is used to normalize the output intensity of the diffractive network's image to reveal the quantitative phase information of the sample in radians, i.e., $I_{QPI}(x,y)$ [rad]. Therefore, at the output plane of the diffractive QPI network, we defined an output signal area that is slightly larger than the input sample field-of-view, where the edges are used to reveal the intensity normalization factor, which makes our diffractive QPI designs *invariant* to changes in the illumination beam intensity or the diffraction efficiency of the imaging system, correctly revealing $I_{QPI}(x,y)$, matching the quantitative phase information of the input object in radians.

Figure 2a shows the phase-only diffractive layers constituting a diffractive QPI network that is trained using deep learning. In our proof-of-concept numerical experiments, we opted to train and test our diffractive network designs on well-known image datasets to better benchmark the resulting QPI capabilities. Given a normalized greyscale image from a target dataset, $\phi(x,y)$, the corresponding function of a phase object at the input plane can be written as $e^{j\alpha\pi\phi(x,y)}$ where $|\phi(x,y)| \leq 1$. The parameter $\alpha$ determines the range of the phase shift induced by the input object. The diffractive optical network shown in Fig. 2a was trained based on $\phi(x,y)$ taken from the Tiny-Imagenet dataset [56] and the parameter, $\alpha$, was set to be 1 for both training and testing, i.e., $\alpha_{tr} = \alpha_{test} = 1$. Figure 2b illustrates the QPI signals, $I_{QPI}(x,y)$, for exemplary test samples from the Tiny-imagenet dataset, never seen by the diffractive network in the training phase, along with the corresponding ground truth images, $\phi(x,y)$. We quantified the success of the QPI signal synthesis performed by the presented diffractive network using the Structural Similarity Index Measure (SSIM) [57] and the peak signal-to-noise ratio (PSNR). The diffractive network shown in Fig. 2a provides an SSIM of 0.824±0.050 (mean ± std) and a PSNR of 26.43dB±2.69 over the entire 10K test samples of the Tiny-Imagenet.

Although our diffractive QPI network design can successfully transform the phase information of the samples into quantitative optical intensity information, providing a competitive QPI performance without the need for any digital phase recovery algorithm, one might argue that the underlying phase-to-intensity transformation performed by the diffractive network is data-specific. To shed more light on this, we investigated the generalization capabilities of our diffractive network design by further testing its QPI performance over phase-encoded samples from two completely different image datasets, i.e., CIFAR-10 and Fashion-MNIST, that were not used in the training phase. As shown in Figs. 2c-d, the SSIM and PSNR values achieved by the presented diffractive QPI network for quantitative phase imaging of CIFAR-10 (and Fashion-MNIST) images are 0.917±0.041 (and 0.596±0.116) and 31.98dB±3.15 (and 26.94dB±1.5), respectively. Interestingly, the QPI signal synthesis quality turned out to be higher for CIFAR-10 images compared to the performance of the same diffractive network on the Tiny-Imagenet test samples, even though CIFAR-10 has an entirely different set of objects and spatial features (which were never used during the training phase). This could be partially attributed to the difference in the original size of the Tiny-Imagenet (64×64-pixel) and CIFAR-10 (32×32-pixel) images. Considering that the physical dimensions of the input field-of-view in our network configuration is 42.4λ×42.4λ, the size of the smallest spatial feature becomes $\frac{42.4\lambda}{64} = 0.6625\lambda$ and $2 \times 0.6625\lambda$ for Tiny-Imagenet and CIFAR-10 datasets, respectively; this makes CIFAR-10 test samples relatively easier to image through the diffractive QPI network.

Next, we numerically quantified the smallest resolvable linewidth and the related phase sensitivity of our diffractive QPI network design using binary phase gratings as test objects (see Fig. 3). Such resolution test targets were not used as part of the training, which only included the Tiny-Imagenet dataset. The presented diffractive network performs QPI with diffractive layers of size 106λ×106λ that are placed 40λ apart from each other and the input/output fields-of-view (see Fig. 1). This physical configuration reveals that the numerical aperture (NA) of our diffractive network is $\sin\left(tan^{-1}\left(\frac{106\lambda}{2\times 40\lambda}\right)\right) = \sim 0.8$, which corresponds to a diffraction-limited resolvable linewidth of $0.625\lambda$. Our numerical analysis in Fig. 3a showed that the smallest resolvable linewidth with our diffractive QPI design was ~$0.67\lambda$, when the input gratings were 0-π encoded, closely matching the resolvable feature size determined by the NA of our system; also note that the effective feature size of the training samples from Tiny-Imagenet is $0.6625\lambda$. This analysis means that our training phase was successful in approximating a general-purpose quantitative phase imager despite using relatively lower resolution training images, coming close to the theoretical diffraction limit imposed by the physical structure of the diffractive QPI network.

The input phase contrast is another crucial factor affecting the resolution of QPI achieved by our diffractive network design. To shed more light on this, we numerically tested our diffractive QPI network on binary gratings with two different linewidths, $0.67\lambda$ and $0.75\lambda$, at varying levels of input phase contrast, as shown in Fig. 3b. Based on the resulting diffractive QPI signals illustrated in Figs. 3a-b, the $0.67\lambda$ linewidth grating remains resolvable until the input phase contrast falls below $0.25\pi$. The last column of Fig. 3b suggests that when the contrast parameter ($\alpha_{test}$) is taken to be 0.1, the noise level in the QPI signal generated by the diffractive network increases to a level where the $0.67\lambda$ linewidth grating cannot be resolved anymore. On the other hand, $0.75\lambda$ linewidth grating remains to be partially resolvable despite the noisy background, even at $0$-$0.1\pi$ phase contrast (i.e., $\alpha_{test} = 0.1$).

We also conducted a similar analysis on the effect of the input phase contrast over the quality of QPI performed by the presented diffractive network. By setting the phase contrast parameter $\alpha_{test}$ to 9 different values between 0.1 and 2.0 for all three image datasets (Tiny-Imagenet, CIFAR-10 and Fashion-MNIST), we quantified the resulting SSIM and PSNR values for the reconstructed images at the output plane of the diffractive QPI network. Figures 4a-c illustrate the mean and standard deviations of the SSIM and PSNR metrics as a function of $\alpha_{test}$ for all three image datasets. A close examination of Figs. 4a-c reveals that both SSIM and PSNR peaks at $\alpha_{test} = 1$, which matches the phase encoding range used during the training phase, i.e., $\alpha_{tr} = \alpha_{test} = 1$. To the left of these peaks, where $\alpha_{test} < \alpha_{tr} = 1$, there is a slight degradation in the performance of the presented diffractive QPI network, mainly due to the increasing demand in phase sensitivity at the resulting image, $I_{QPI}(x, y)$. With $\alpha_{tr} = 1$ and 8-bit quantization of input signals, the phase step size that the diffractive QPI network was trained with was $\frac{\pi}{256} = 0.0123$ radians; however, when $\alpha_{test}$ deviates from the training, for instance $\alpha_{test} = 0.5$, then the smallest phase step size that the diffractive network is tasked to sense becomes $\frac{0.5\pi}{256} = 0.0062$ radians. In other words, the diffractive network must be 2× more phase sensitive compared to the level it was trained for, causing some degradation in the SSIM and PSNR values as shown in Figs. 4a-c for $\alpha_{test} < \alpha_{tr} = 1$.

On the other hand, when the input phase encoding exceeds the $[0, \pi]$ range used during the training phase, the degradation in diffractive QPI signal quality is more severe. As $\alpha_{test}$ approaches to 2.0, the errors and artifacts created by the presented diffractive network in computing the QPI signal increase. Interestingly, at $\alpha_{test} = 1.99$, the forward optical transformation of the diffractive QPI network starts to act as an edge detector. A straightforward solution to mitigate this performance degradation is to train the diffractive network with $\alpha_{tr} = 2.0 - \epsilon$, where $\epsilon$ is a small number, meaning that during the training phase, the dynamic range of the phase values at the input plane will be within $[0, 2\pi)$. Supplementary Fig. S1 illustrates an example of this for a 5-layer diffractive QPI network that was trained with $\alpha_{tr} = 1.99$. This new diffractive network has the same physical layout and architecture as the previous one shown in Fig. 2. The only difference between the two diffractive QPI networks is the phase range covered by the input samples used during their training ($\alpha_{tr} = 1.0$ vs. $\alpha_{tr} = 1.99$). Since the design evolution of this new diffractive QPI network is driven by input samples covering the entire $[0, 2\pi)$ phase range, in the case of $\alpha_{test} = \alpha_{tr} = 1.99$, it provides a much better QPI performance compared to the diffractive network shown in Fig. 2. This improved diffractive QPI performance can also be visually observed by comparing the images shown in Fig. 4 and Supplementary Fig. S1 under the $\alpha_{test} = 1.99$ column.

**Discussion**

A vital feature of the presented diffractive QPI networks is that their operation is invariant to changes in the input beam intensity or the power efficiency of the diffractive detection system; by using the mean intensity value surrounding the output image field-of-view as a normalization factor, the resulting diffractive image intensity $I_{QPI}(x, y)$ reports the phase distribution of the input object in radians. Moreover, the presented diffractive optical networks are composed of passive layers, and therefore perform QPI without any external power source other than the illumination light. It is true that the training stage of a diffractive QPI network takes a significant amount of time (e.g., ~40 hours) and consumes some energy for training-related computing. But this is a one-time training effort, and in the image inference stage, there is no power consumption per object (except for the illumination), and the reconstructed image reveals the quantitative phase information of the object at the speed of light propagation through a passive network, without the need for a graphics processing unit (GPU) or a computer. One should think of a diffractive network's design, training and fabrication phase (a one-time effort) similar to the design/fabrication/assembly phase of a digital processor or a GPU that we use in our computers.

The output power efficiency of the presented QPI networks is mainly affected by two factors: diffraction efficiency of the resulting network and material absorption. In this study, we assumed the optical material of diffractive surfaces has a negligible loss for the wavelength of operation, similar to the properties of optical glasses, e.g., BK-7, in the visible part of the spectrum. For example, the diffractive QPI network reported in Fig. 2 achieves ~2.9% mean diffraction efficiency for the entire 10K test set of Tiny-Imagenet. It is important to note that during the training of this diffractive QPI network, the training cost/loss function was purely based on decreasing the QPI errors at the output plane, and there was no other loss term or regularizer to enforce a more power-efficient operation. In fact, by including an additional loss term for regulating the balance between the QPI performance and diffraction efficiency (see Methods section), we demonstrated that it is possible to design more efficient diffractive QPI networks with a minimal compromise on the output image quality; see Fig. 5, where all the diffractive network designs share the same physical layout shown in Fig. 1. For example, a more efficient diffractive QPI network design with 6.31% power efficiency at the output plane offers QPI signal quality with an SSIM of 0.815±0.0491. Compared to the original diffractive QPI network design that solely focuses on output image quality, the SSIM value of this new diffractive network has a negligible decrease while its diffraction efficiency at the output plane is improved by more than 2-fold. Further shifting the focus of the QPI network training towards improved power efficiency can result in a solution that can synthesize QPI signals with >11% output diffraction efficiency, also achieving an SSIM of 0.771±0.0507 (see Fig. 5).

Another crucial parameter in a diffractive network design is the number of diffractive layers within the system; Supplementary Figure S2 illustrates the results of our analysis on the relationship between the diffractive QPI performance and the number of diffractive layers within the system. It has previously been shown through both theoretical and empirical evidence that deeper diffractive optical networks can compute an arbitrary complex-valued linear transformation with lower approximation errors, and they demonstrate higher generalization capacity for all-optical statistical inference tasks [58,59]. Supplementary Figure S2 confirms the same behavior: improved QPI performance is achieved by increasing the number of diffractive layers, $K$. When $K=1$, the trained diffractive network fails to compute the QPI signal for a given input phase object, as evident from the extremely low SSIM values and the exemplary images shown in Fig. S2b. On top of that, the diffraction efficiency is also very low, ~1%, with a single-layer diffractive network configuration ($K=1$). With $K=2$ trainable diffractive surfaces, the diffraction efficiency stays very low, while the QPI signal quality improves. When we have $K=3$ diffractive layers in our QPI network design, we observe a significant improvement in both the diffraction efficiency and the output SSIM compared to *K*=1 or 2. Beyond *K*=3, the structural quality of the output QPI signal keeps improving as we add more layers to the diffractive network architecture. However, this improvement does not translate into better diffraction efficiency as the training loss function does not include a power efficiency penalty term. Earlier results reported in Fig. 5 clearly show the impact of adding such a regularizer term in the training loss function for improving the diffraction efficiency of the QPI network, reaching >11% power efficiency with a minor sacrifice in the structural fidelity of the output images.

In summary, the presented diffractive QPI networks convert the phase information of an input object into an intensity distribution at the output plane in a way that the normalized output intensity reveals the phase distribution of the object in radians. Being resilient to input light intensity variations and power efficiency changes in the diffractive set-up, this QPI network can replace the bulky lens-based optical instrumentation and the computationally intensive reconstruction algorithms employed in QPI systems, potentially offering high-throughput, low-latency, compact and power-efficient QPI platforms which might fuel new applications in on-chip microscopy and sensing. Fabrication and assembly of such diffractive QPI systems operating in the visible and near IR wavelengths can be achieved using two-photon polymerization-based 3D printing methods as well as optical lithography tools [60–62].

**Methods**

**Optical Forward Model of Diffractive QPI Networks**

The optical wave propagation in air, between successive diffractive layers, was formulated based on the Rayleigh-Sommerfeld diffraction equation. According to this formulation, the free-space propagation inside a homogeneous and isotropic medium is modeled as a shift-invariant linear system with the impulse response,

$$w(x,y,z) = \frac{z}{r^2}\left(\frac{1}{2\pi r} + \frac{n}{j\lambda}\right)\exp(\frac{j2\pi n r}{\lambda}) \qquad 1$$

where $r = \sqrt{x^2 + y^2 + z^2}$. In Eq. 1, the parameters $n$ and $\lambda$ denote the refractive index of the medium ($n = 1$ for air), and the wavelength of the illumination light, respectively. Accordingly, a diffractive neuron, $i$, located at $(x_i, y_i, z_i)$ on $k^{th}$ layer can be considered as the source of a secondary wave, $u_i^k(x, y, z)$,

$$u_i^k(x, y, z) = w_i(x, y, z) t(x_i, y_i, z_i) \sum_{q=1}^{N} u_q^{k-1}(x_i, y_i, z_i) \qquad 2.$$

where the summation in Eq. 2 represents the field generated over the diffractive neuron located at $(x_i, y_i, z_i)$ by the neurons on the previous, $(k-1)^{th}$, layer. From Eq. 1, the function $w_i(x, y, z)$ in Eq. 2 can be written as,

$$w_i(x, y, z) = \frac{z - z_i}{r^2} \left( \frac{1}{2\pi r} + \frac{n}{j\lambda} \right) \exp(\frac{j 2\pi n r}{\lambda}) \qquad 3,$$

with $r = \sqrt{(x - x_i)^2 + (y - y_i)^2 + (z - z_i)^2}$. The multiplicative term $t(x_i, y_i, z_i)$ in Eq. 2 denotes the transmittance coefficient of the neuron, $i$, which, in its general form, can be written as, $t(x_i, y_i, z_i) = a_i \exp(j\theta_i)$. Depending on the diffractive layer fabrication method and the related optical materials, both $a_i$ and $\theta_i$ might be a function of other physical parameters, e.g., material thickness in 3D printed diffractive layers and driving voltage levels in spatial light modulators. In earlier works on diffractive networks [53,54,63,64], it has been shown that it is possible to directly train such physical parameters through deep learning. On the other hand, a more generic way of optimizing a diffractive network is to define the amplitude $a_i$ and $\theta_i$ as learnable parameters. In this study, we constrained our analysis to phase-only diffractive surfaces where the amplitude coefficients, $a_i$, were all taken as 1 during the entire training. Thus, the only learnable parameters of the presented diffractive networks are the phase shifts applied by the diffractive features, $\theta_i$. For all the diffractive networks that we trained, the initial value of all $\theta_i$s was set to be 0, i.e., the initial state of a diffractive network (before the training kicks in) is equal to the free-space propagation of the input light field onto the output plane.

**The design of diffractive QPI networks**

During our deep learning-based diffractive network training, we sampled the 2D space with a period of $0.53\lambda$, which is also equal to the size of each diffractive feature ('neuron') on the diffractive surfaces. Although we described the forward optical model over continuous functions in the previous subsection, training of the presented diffractive networks was performed using digital computers. Hence, we denote the input and output signals using their discrete counterparts for the remaining part of this sub-section with a spatial sampling period of 0.53λ in both directions (*x* and *y*). In the physical layout of the presented diffractive optical networks, the size of the input field-of-view was set to be 42.4λ×42.4λ, which corresponds to 80×80 2D vectors defining the phase distributions of input objects. With $I[m,n]$ denoting an image of size $M \times N$ from a dataset, we applied 2D linear interpolation to compute the 2D vector $\boldsymbol{\phi}[q,p]$ of size 80×80. Note that the values of $M$ and $N$ depend on the used image dataset. Specifically, for Tiny-Imagenet $M = N = 64$, while for CIFAR-10 and Fashion-MNIST datasets, $M = N = 32$ and $M = N = 28$, respectively. The scattering function within the input field-of-view of the diffractive networks was defined as a pure phase function (see Fig. 1) in the form of $e^{j\alpha\pi\boldsymbol{\phi}[q,p]}$.

The physical dimensions of each diffractive layer were set to be 106λ on both *x* and *y* axes, i.e., each diffractive layer contains 200×200 = 40K neurons. For instance, the 5-layer diffractive network shown in Fig. 2 has 0.2 million neurons, and hence 0.2 million trainable parameters, $\theta_i$, $i = 1,2,\ldots,0.2 \times 10^6$. In our forward optical model, we set all the distances between (1) the first diffractive layer and the input field-of-view, (2) two successive diffractive layers, and (3) the last diffractive layer and the output plane, as 40λ resulting in an NA of ~0.8. With the size of each diffractive feature/neuron taken as 0.53λ, the diffraction cone angle of the secondary wave emanating from each neuron ensures optical communication between all the neurons on two successive surfaces (axially separated by 40λ), while also enabling a highly compact diffractive QPI network design. For instance, the total axial distance from the input field-of-view to the output plane of a 5-layer diffractive QPI network shown in Fig. 1 is only ~240λ.

The size of the QPI signal area at the output plane including the reference/background region was set to be 43.56λ×43.56λ, i.e., the reference region extends on both directions on *x* and *y* axes by 0.53λ, (43.56λ=42.4λ+2×0.53λ). If we denote the background optical intensity over this reference region as $\boldsymbol{I_R}[r]$ and the optical intensity within the QPI signal region as $\boldsymbol{I_S}[q,p]$, then according to our forward model, $\boldsymbol{I_{QPI}}[q,p]$ is found by,

$$\boldsymbol{I_{QPI}}[q,p] = \frac{\boldsymbol{I_S}[q,p]}{B}, \qquad 4$$

where $B = \frac{1}{N_R}\sum_{r=1}^{N_R} I_R[r]$ is the mean background intensity value, $N_R$ denotes the number of discretized intensity samples within the reference region. According to Eq. 4, for a given input object/sample, the final diffractive QPI signal, $I_{QPI}[q,p]$ reports the output phase image in radians.

To guide the evolution of the diffractive layers according to the QPI signal in Eq. 4, at each iteration of the deep learning-based training of the presented diffractive QPI networks, we updated the phase parameters, $\theta_i$, using the following normalized mean-squared-error [65],

$$\mathcal{L} = \frac{1}{N_R + N_S} \sum_{l=1}^{N_R+N_S} |o[l] - \sigma o'[l]|^2, \qquad 5$$

where, $N_s$ is the total number of discretized samples representing the QPI signal area, i.e., $N_s = 80 \times 80$. The vectors $o$ and $o'$ are 1D counterparts of the associated 2D discrete signals, $o[q,p]$ and $o'[q,p]$, computed based on lexicographically ordered vectorization operator. They denote the ground-truth phase signal of the input object and the diffractive intensity signal synthesized by the QPI network at a given iteration, respectively. Both the ground truth vector, $o$, and $o'$ cover the output sample field-of-view and the reference signal region surrounding it, hence their size is equal to $N_R + N_S = 82 \times 82$. The 2D vector $o[q,p]$ is defined based on the input vector $\phi[q,p]$. First, we equalize the size of the two vectors by padding the $80 \times 80$ vector $\phi[q,p]$ to the size $82 \times 82$. The values over the padded region are equal to $\frac{1}{\alpha\pi}$. This padded vector was then scaled with the multiplicative constant $\alpha\pi$ such that the $80 \times 80$ part in the middle represents the argument of the phase function $e^{j\alpha\pi\phi[q,p]}$. The reference signal region surrounding this $80 \times 80$ part has all ones, implying that the mean intensity over this area will correspond to 1 rad. By computing the loss function in Eq. 5 based on a ground-truth vector that also includes the desired reference signal intensity, we implicitly enforce/train the diffractive QPI network to synthesize a uniformly distributed intensity over the reference signal area, although this is not a requirement for the QPI networks' operation.

The multiplicative term, $\sigma$, in Eq. 5 is a normalization constant that was defined as [65],

$$\sigma = \frac{\frac{1}{N_R + N_S}\sum_{l=1}^{N_R+N_S} o[l]o'^*[l]}{\frac{1}{N_R + N_S}\sum_{l=1}^{N_R+N_S}|o'[l]|^2}, \qquad 6$$

The structural loss function, $\mathcal{L}$, in Eq. 5 drives the QPI quality, and it was the only loss term used during the training of the diffractive networks shown in Fig. 2, Supplementary Figs. S1 and S2. The training of the diffractive network designs with output diffraction efficiencies of $\geq 2.9\%$ shown in Fig. 5, on the other hand, use a linear mix of the structural loss in Eq. 5 and an additional loss term penalizing poor power efficiency, i.e., $\mathcal{L}' = \mathcal{L} + \gamma \mathcal{L}_p$. The functional form of the power efficiency-related penalty $\mathcal{L}_p$ was defined as,

$$\mathcal{L}_p = e^{-\eta}, \qquad 7$$

where $\eta$ stands for the percentage of power efficiency,

$$\eta = \frac{P_{out}}{P_1} \times 100, \qquad 8$$

with $P_1$ denoting the optical power incident on the 1st diffractive layer and $P_{out} = \sum_{l=1}^{N_R+N_S} |\boldsymbol{o}'[l]|^2$. The coefficient $\gamma$ is a multiplicative constant that determines the weight of the power efficiency-related term in the total loss, $\mathcal{L}'$. The value of $\gamma$ directly affects the diffraction efficiency of the resulting diffractive QPI network design. Specifically, for the diffractive network shown in Fig. 2, it was set to be 0. On the other hand, when $\gamma$ was taken as 0.1, 0.4 and 5.0, the corresponding diffractive QPI network designs achieved 6.31%, 8.17% and 11.05% diffraction efficiency ($\eta$), respectively (see Fig. 5).

**Implementation Details of Diffractive QPI Network Training**

The deep learning-based diffractive QPI network training was implemented in Python (v3.7.7) and TensorFlow (v1.15.0, Google Inc.). For the gradient-based optimization, we used the Adam optimizer with its momentum parameter $\beta_1$ set to 0.5 [66]. The learning rate was taken as 0.01 for all the presented diffractive QPI networks. With the batch size equal to 75, we trained all the diffractive networks for 200 epochs, which takes ~40 hours using a computer with a GeForce GTX 1080 Ti GPU (Nvidia Inc.) and Intel® Core ™ i7-8700 Central Processing Unit (CPU, Intel Inc.) with 64 GB of RAM, running Windows 10 operating system (Microsoft). To avoid any aliasing in the representation of the free-space impulse response depicted in Eq. 1, the dimensions of the simulation window were taken as 1024×1024.

The PSNR image metric was calculated as follows:

$$PSNR = 20 log_{10} \left( \frac{\alpha\pi}{\sqrt{\left|\alpha\pi\boldsymbol{\phi}[q,p] - \boldsymbol{I_{QPI}}[q,p]\right|^2}} \right), \qquad 9$$

For SSIM calculations, we used the built-in function in Tensorflow, i.e., tf.image.ssim, where the two inputs were $\alpha\pi\boldsymbol{\phi}[q,p]$ and $\boldsymbol{I_{QPI}}[q,p]$, representing the ground-truth image and the QPI signal synthesized by the diffractive network, respectively. The input parameter "max_val" was set to be $\alpha\pi$ in these SSIM calculations.

**Figures**

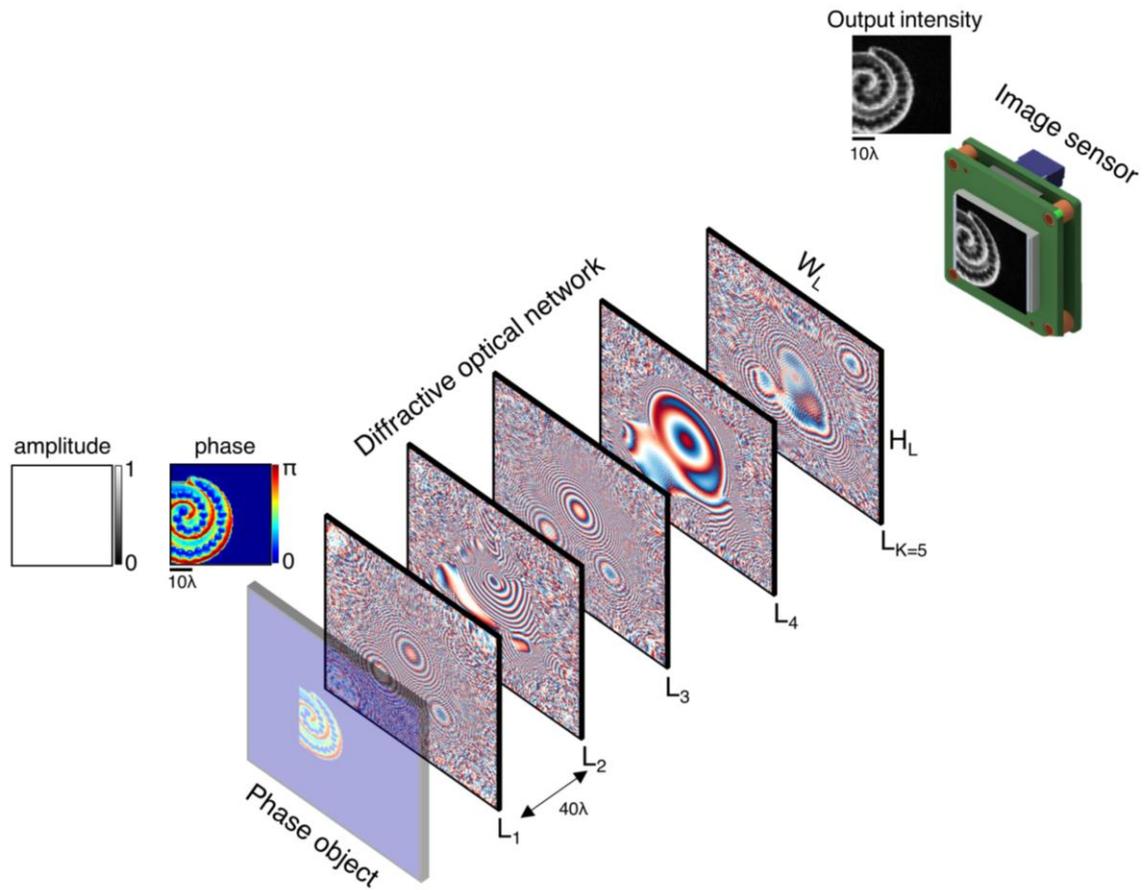

**Figure 1. Schematic of a diffractive QPI network that converts the optical phase information of an input object into a normalized intensity image, revealing the QPI information in radians without the use of a computer or a digital image reconstruction algorithm.** Optical layout of the presented 5-layer diffractive QPI network, where the total distance between the input and output fields-of-view is 240λ.

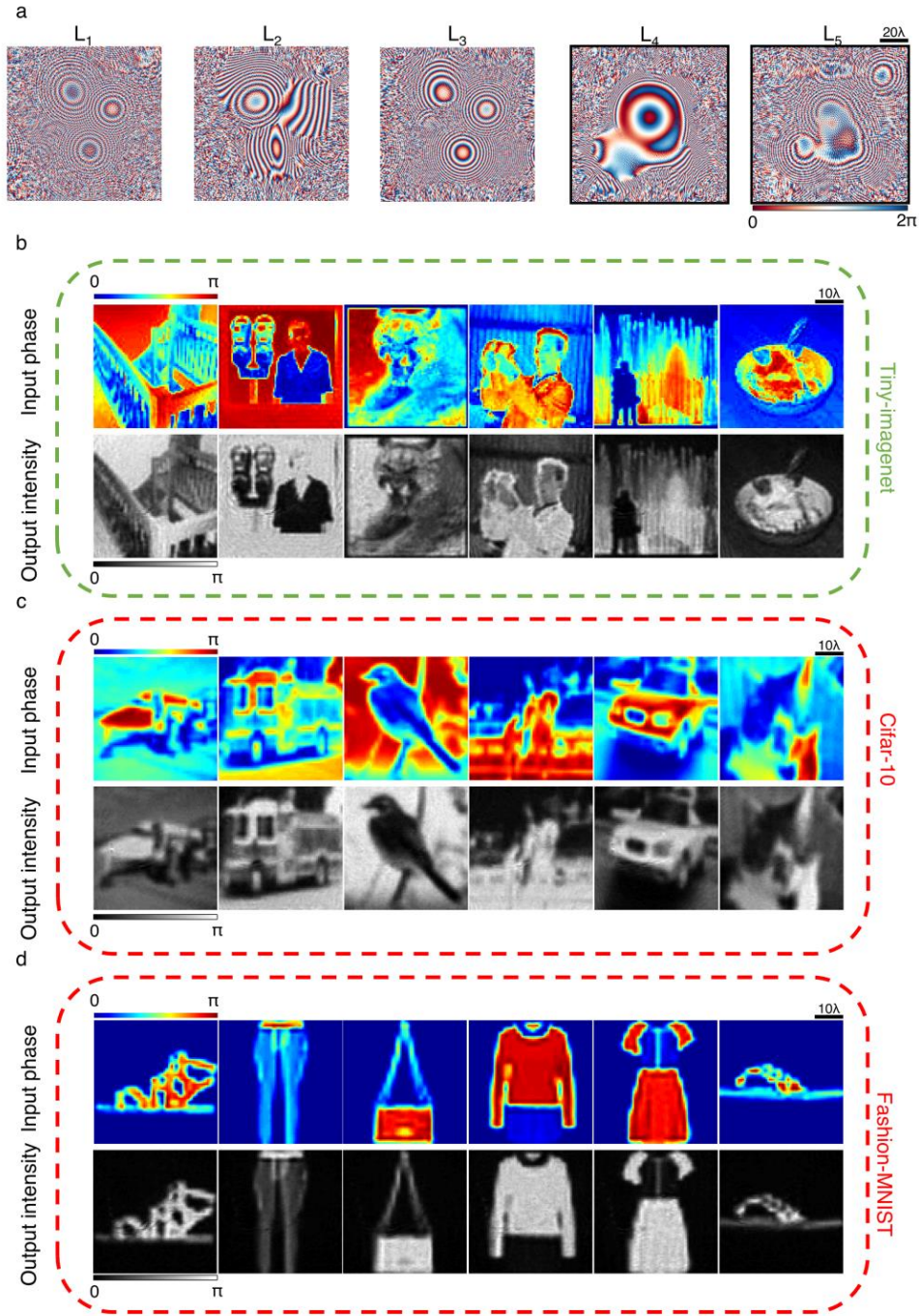

**Figure 2. Generalization capability of diffractive QPI networks. a**, The phase profiles of the diffractive layers forming the diffractive QPI network trained using phase-encoded images from Tiny-Imagenet dataset, $\phi(x, y)$. **b**, Exemplary input object images and the corresponding output QPI signals for the test images, never seen by the network during training, taken from the Tiny-Imagenet. Dashed green box indicates that the test images, although not seen by the diffractive network before, belong to the same dataset used in the training. **c-d**, Same as **b**, except that the test images are taken from CIFAR-10 and Fashion-MNIST. Dashed red boxes indicate that these test images are from entirely new datasets compared to the Tiny-Imagenet used in the training. The SSIM (PSNR) values achieved by the presented diffractive network are 0.824±0.050 (26.43dB±2.69), 0.917±0.041 (31.98dB±3.15) and 0.596±0.116 (26.94dB±1.5) for the test images from Tiny-Imagenet, CIFAR-10 and Fashion-MNIST datasets, respectively.

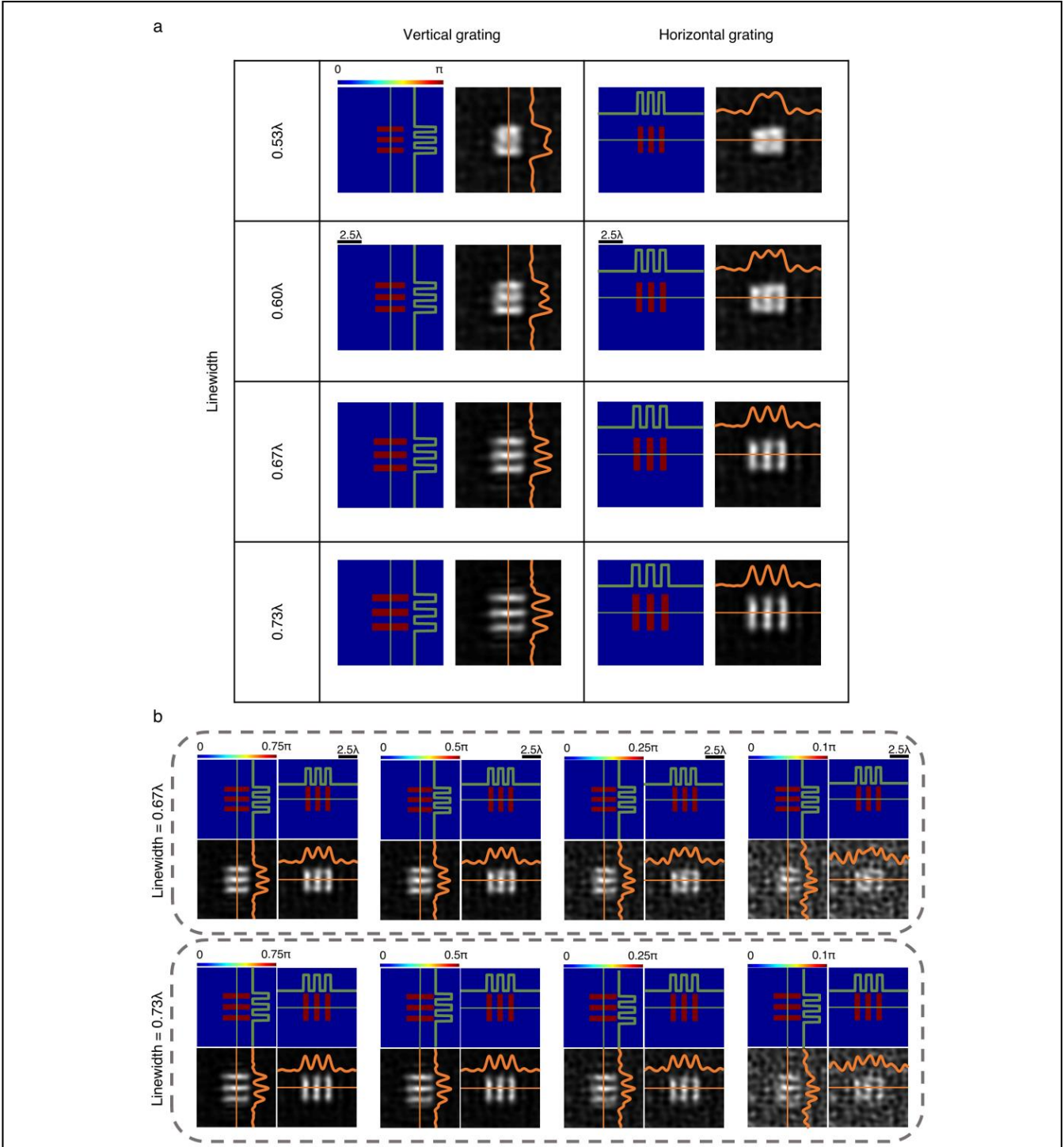

**Figure 3. Spatial resolution and phase sensitivity analysis for the diffractive QPI network shown in Fig. 2. a**, Input phase image and the corresponding output diffractive QPI signal for binary, 0-π phase encoded grating objects. The diffractive QPI network can resolve features as small as ~0.67λ. **b**, Analysis of the relationship between the input phase contrast and the resolvable feature size. The diffractive QPI network can resolve 0.67λ linewidth for a phase encoding range that is larger than 0.25π. Below this phase contrast, the resolution slowly degrades; for example, at 0-0.1π phase encoding, the background noise shadows the QPI signal of the grating with a linewidth of 0.67λ, while a larger linewidth (0.73λ) grating is still partially resolvable.

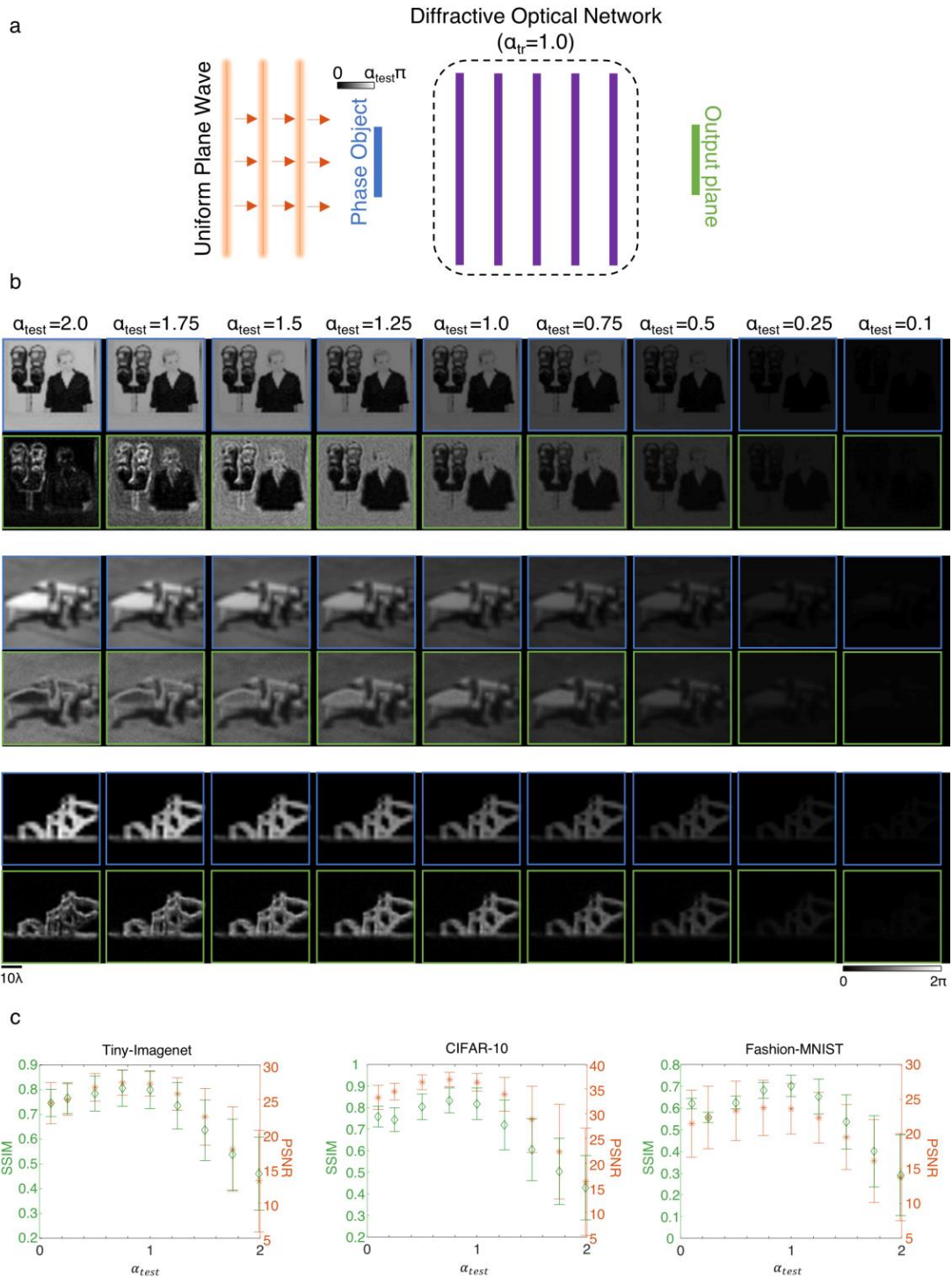

**Figure 4. The impact of input phase range on the diffractive QPI signal quality. a**, A schematic of the diffractive QPI network that was trained with $\alpha_{tr} = 1.0$, meaning that the training images had $[0 : \pi]$ phase range. **b**, Pairs of ground-truth input phase images (top rows) and the diffractive QPI signal (bottom rows) for different images taken from Tiny-Imagenet (top), CIFAR-10 (middle) and Fashion-MNIST (bottom), at different levels of phase encoding ranges dictated by (from left-to-right) $\alpha_{test} = 2$, $\alpha_{test} = 1.75$, $\alpha_{test} = 1.5$, $\alpha_{test} = 1.25$, $\alpha_{test} = \alpha_{tr} = 1.0$, $\alpha_{test} = 0.75$, $\alpha_{test} = 0.5$, $\alpha_{test} = 0.25$, $\alpha_{test} = 0.1$. **c**, The SSIM and PSNR values of the diffractive QPI signals with respect to the ground-truth images as a function of $\alpha_{test}$.

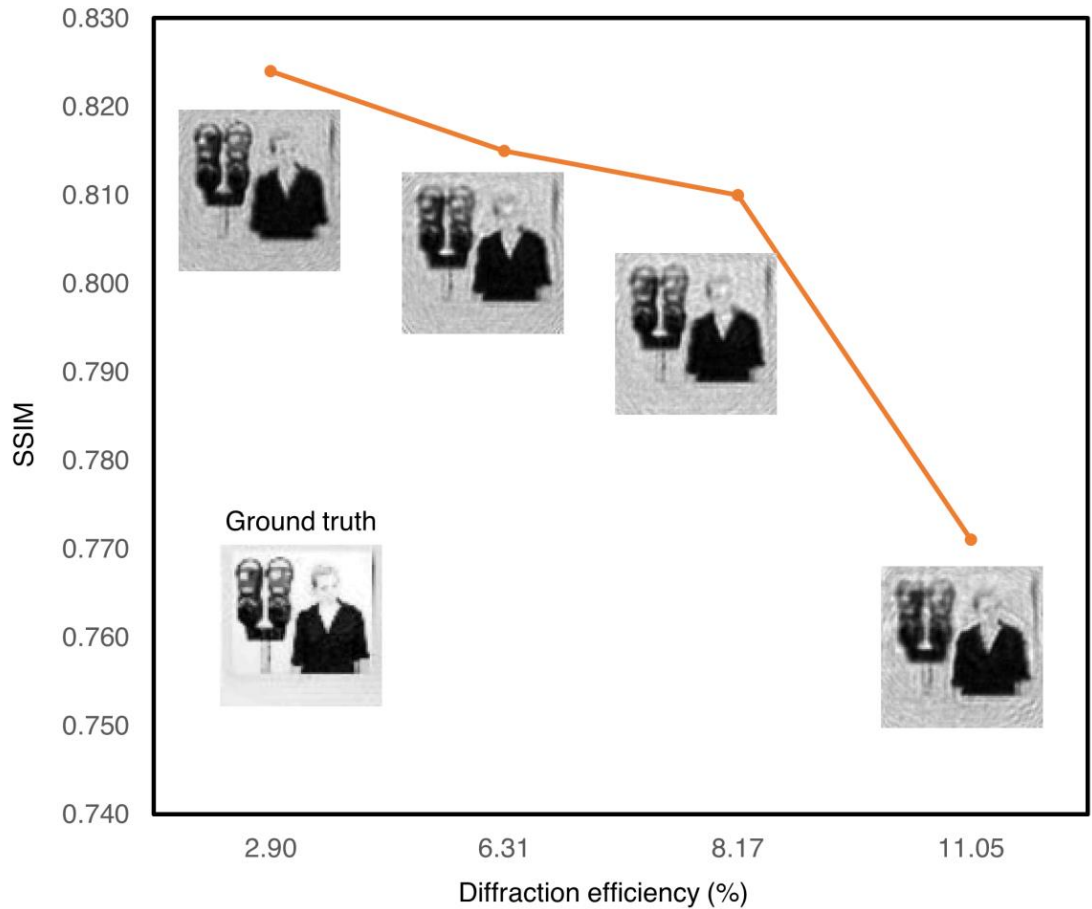

**Figure 5. Diffractive QPI signal quality and the power efficiency trade-off.** We report 4 different diffractive QPI network models trained using $[0 : \pi]$ phase-encoded samples from the Tiny-Imagenet dataset. The SSIM on the y-axis reflects the mean value computed over the entire 10K test images of the Tiny-Imagenet dataset. The diffractive QPI network that provides the highest SSIM is the network shown in Fig. 2, which was trained solely based on the structural loss function (Eq. 5) totally ignoring the diffraction efficiency of the resulting solution. The loss function used for the training of the other 3 diffractive QPI networks includes a linear superposition of the structural loss function (Eq. 5) and the diffraction efficiency penalty term depicted in Eq. 7. The multiplicative constant $\gamma$ which determines the weight of the diffraction efficiency penalty was taken as 0.1, 0.4 and 5.0 for these 3 diffractive QPI networks, providing an output diffraction efficiency of 6.31% , 8.17% and 11.05%, respectively.